# Proposed platform for improving grid security by trust management system

Safieh Siadat
Islamic Azad University,
Science and Research Branch,
Tehran, Iran

Amir Masoud Rahmani
Islamic Azad University,
Science and Research Branch,
Tehran, Iran

Mehran Mohsenzadeh
Islamic Azad University,
Science and Research Branch,
Tehran, Iran

*Abstract*— With increasing the applications of grid system, the risk in security field is enhancing too. Recently Trust management system has been recognized as a noticeable approach in enhancing of security in grid systems. In this article due to improve the grid security a new trust management system with two levels is proposed. The benefits of this platform are adding new domain in grid system, selecting one service provider which has closest adaption with user requests and using from domains security attribute as an important factor in computing the trust value.

## I. INTRODUCTION

Grid computing is a newly developed technology for complex systems with large-scale resource sharing, wide-area communication, and multi-institutional collaboration [1]. Due to the complexity of grid computing, the traditional network security practices cannot meet the security requirement of grid. As a result, trust management is crucial to security and trustworthiness in grids. Security and trust are two distinct concepts. In literature trust has been sometimes termed as "soft security" and can implement sophisticated security decisions. So the TMS will not replace GSI, it only assist it to provide more refined and rational choices for Grid security [2]. In this paper a novel TMS with tow levels is proposed. The goal of this platform is optimizing available TMS in the grid systems. Our TMS is a comprehensive platform in grid environment and try to remove the weakness of the old platform. In new platform there are components such as security management and demand trust evaluation that old trust management system in grid environment had not paid attention yet. The presence of these components is crucial in making true decision. Security management component has used for measuring of different domain security level in grid systems. Demand trust evaluation component select one service provider which has closest adaption with user requests.

Other component used in this platform are trust negotiation, registration, propagation, feedback evaluation, trust evaluation, access control and monitoring which they have specified task. Trust negotiation task is to add of new domain in grid systems. The activity of registration component is to register new domain properties in grid systems. Propagation component task is to broadcast of new domain properties for all domains in grid systems. Feedback evaluation component duty is evaluate and update received feedback from service requester. Trust evaluation component task is to compute servers trust value based on received feedback, user satisfaction value and self defense capability in each domain. The task of accesses control component is accessing control on available repository and the duty of monitoring component is trust re-evaluation and adding new information in TMS.

*Outline of the paper:* In Section 2, related work is presented. Section 3 propose newly developed platform. At last a conclusion and future work is given in Section 4.

## II. RELATED WORK

Trust management was first introduced by Blaze, et al. in 1996 [3], and many trust management models were proposed, for instance, PolicyMaker [3], KeyNote [4], REFEREE [5], SPKI/SDSI [6]. Recently trust management is known as a new method to make secure grid systems and some researches is done using TMS in grid systems. A number of researches are mentioned below.

The problems of managing trust in Grid environments are discussed by Azzedin and Maheswaran [7]-[9]. They define the notion of trust as consisting of identity trust and behavior trust. They separate the "Grid domain" into a "Client domain" and a "resource domain", and the way they calculate trust is limited in terms of computational scalability, because they try to consider all domains in the network; as the number of domains grows, the computational overhead grows as well. Hwang et al. [10] and Sobolewski [11] try to build trust and security models for Grid environments, using trust metrics based on e-business criteria. Alunkal et al. [12] propose to build an infrastructure called "Grid Eigentrust" using a hierarchical model in which entities are connected to institutions which then form a VO. They conclude with the realization of a "Reputation Service", however, without providing mechanisms that automatically can update trust values. Papalilo and Freisleben [13] has proposed a Bayesian based Trust model for Grid but the suggested metrics cover only limited trust aspects in practical Grid. TieYan et al. [14] consider trust only to improve the Grid Security Infrastructure (GSI) to achieve additional authentication means between Grid users and Grid services. Ching et al. [16] use the concepts of the subjective logic in the context of Grid computing using trust relationships to enhance the grid security. M.H. Durad, Y. Cao proposed grid trust management system. In their research only the platform was described, while there was not the comprehensive description of components mathematically [2]. In this article to conquer above problem a complete platform including mathematic formulation is proposed.

### III. PROPOSED PLATFORM

As shown in Fig. 1 the proposed platform has two levels that in next section will be explained. In newly developed platform there is one DTM[1] in each domain of grid system that its task is managing the available resource nodes in that domain. DTM is one of resource nodes in every domains selected by using Ring algorithm. Also there is one GRM[2] that its task is managing DTMs. GRM is one of DTMs selected by Ring algorithm and located in upper level of platform. In order to increasing in fault tolerance, there are back up of DTM and GRM.

*A. Upper level of platform*

There is GRM in upper level which its task is management of DTM. In this level there is virtual mapping of DTM from different domains. By this way the neighborhood of domains will be saved in grid systems. Upper level includes 3 components:
1-Trust negotiation component
2- Registration and initialization component
3- Propagation component.

*1) Trust negotiation component*

The task of this component is adding new domain in grid systems. The trust negotiation component has two levels:
  a) Authentication level.
  b) Policy mapping level.
This component is illustrated in Fig.2.

  *a) Authentication level*

This level accomplishes the authentication of new domain that wants to be added in grid systems.

  *b) Policy mapping level*

The task of this level is to adapt the policy of grid domains with the new domain policy. After adaption process if there is minimum satisfaction between new domain and grid domains, new domains will be authorized for adding in grid systems.

DEFINITION 1. MINIMUM SATISFACTION

As shown in relation (1) and (2), If $c_1, c_2, c_3, ....c_k$ are defined and agreed policy in grid system, the domain has authorized to be added into grid system which can satisfy half of $c_1, c_2, c_3, ...c_k$ at least.

$$c_1, c_2, c_3, ......, c_k \in C. \quad (1)$$

$$c'_1, c'_2, c'_3, .....c'_l \subset C \ , \ l \geq k/2. \quad (2)$$

Fig.3 illustrates algorithm of adding new domain in grid systems.

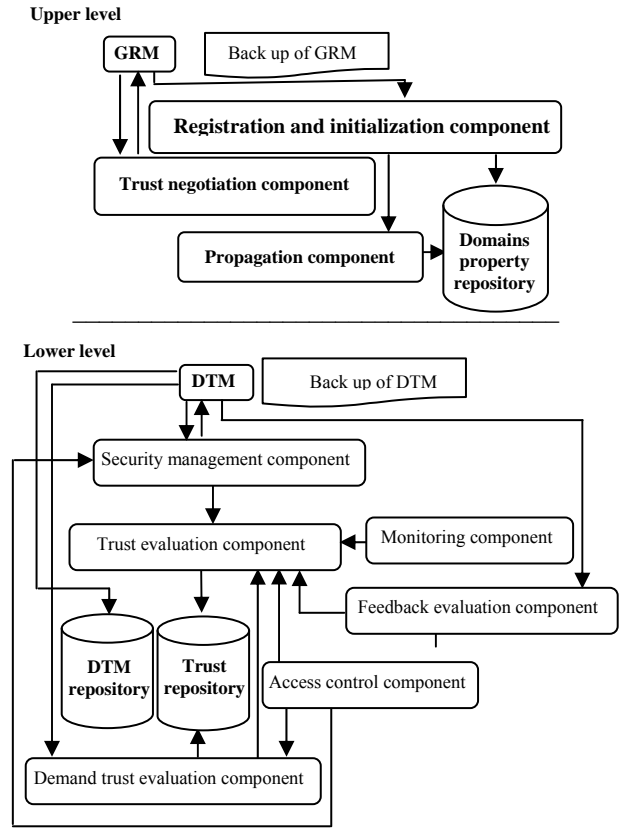

Figure 1. Proposed platform

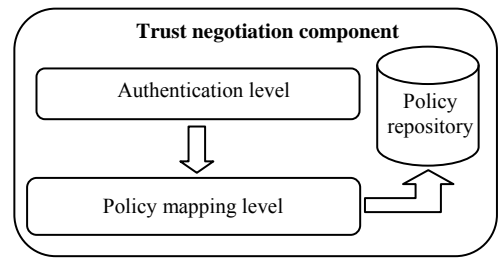

Figure 2. Trust negotiation component

*2) Registration and initialization component*

The activity of this component is to register new domain property which is authorized by trust negotiation component in domain property repository. The other task of registration and initialization component is initializing trust value of new domain resource node by 0.5 because in this platform has been assumed that to every resource nodes added into grid system, the allocated trust value is middle.

---
[1] Domain Trust manager (**DTM**)
[2] Global Resource manager (**GRM**)

```
1 Begin
2 new domains send to GRM adding request to grid system;
3 GRM call trust negotiation component ( );
4 if (trust negotiation component authorize new domain) then
5        Goto 8;
6 Else
7        Goto 11;
8 New Domain send to GRM new domain properties;
9 GRM call registration and initialization component ( );
10 registration and initialization component call
```

Figure 3. adding new domain in grid systems

*3) Propagation component*

The task of propagation component is broadcasting new domain properties for all domains in grid systems.

### B. Lower level of platform

This level includes domains in grid system. There is one DTM in each domain that its task is management of resource nodes. Lower level includes 6 components:

1- Security management component
2- Feedback evaluation component
3- Demand trust evaluation component
4- Trust evaluation component
5- Accesses control component
6- Monitoring component.

Fig. 4 shows lower level of platform algorithm.

```
1 Begin
2 DTM receives request (C, D, PL, Q, T)
// C ∈ (service-request, feedback, security)
// D ∈ (inter-domain, intra-domain)
//PL ∈ (parameter-list)
// Q = DTM-number or resource-node number
// T= type of service
3 DTM sends request for security management component ()
4 if (security management component() authorized request) then
5    goto 8
6 else
7    goto 18
8 security management component sends request to DTM
9 DTM checks C in request
10 if (C = service request) then
11      DTM calls demand trust evaluation (PL, Q)
12 else if (C=feedback) then
13      DTM calls feedback evaluation component (PL)
14 else if (C = security and D = intra-Domain)
15      DTM calls security management component ()
16 else
17      goto 18
18 trust evaluation component ( )
18 End.
```

Figure 4. Lower level of platforme algorithm

*1) Security management component*

This component has been used for measuring different domain security level in grid systems. In this platform the domain security level has been applied as important factor for measuring resource node trust value in each domain. As shown in Fig.5 this component has two levels:

a) Authentication level
b) Self defense capability level

*a) Authentication level*

The received request to each domain will be authenticated by accessing the certificate repository in this level also registering DTM certificate property of each domain in its certificate repository is the task of this level. Two mentioned jobs will be done by authorization and DTM registry management.

*a) Self defense capability*

This level task is to evaluate the self defense capability of different domains in grid system. The self defense ability of different grid domains will be calculated by using of security attribute. Security attributes and evaluation criteria of theirs are shown in table 1. Relation (3) calculate self defense capability different domain in grid systems where as $Sa_i$ is security attribute and $w_i$ is weight of each security attribute.

$$DF(new) = \sum_{i=1}^{m} w_i \times Sa_i. \quad (3)$$

*2) Feedback evaluation component*

This component duty is evaluate and update received feedback from service requester after receiving service. Feedback is a statement issued by a client about the quality of a service or product provided by the service provider after transaction. As shown in Fig.6 Feedback evaluation component has 3 levels:

a) Feedback collection level
b) Feedback verification level
c) Feedback updating level

*a) Feedback collection level*

This level has been used for collecting received feedback and sending them to feedback verification level.

*b) Feedback verification level*

The task of this level is investigating the received feedback by below sub process:

1- identification
2- legitimacy
3- Reasonability
4- Time
5- Rectification

Above sub processes are described in [15]. The only change is on reasonability sub processes. The modification in reasonability sub process is shown in Fig.7 that $f_{pi}(new)$ represent received feedback of *ith* parameter and *a* is average of the end *l* feedbacks.

*a) Feedback updating level*

The duty of this level is updating received feedback from feedback verification level in feedback repository.

*3) Demand trust evaluation component*

This component receives the user requests according to getting a service. Based on user request the best server will be selected for providing the service. To achieve this aim users initialize service quality parameters determined in the platform. It should be noticed that user enters the service quality parameters based on percentage. This component selects the server which has nearest adaption with the request of user. Demand trust evaluation component respond to the user request in batch manner. This component includes two levels:

a) Trust evaluation with demand parameter level
b) Server selection and request allocation.

This component is illustrated in Fig.8.

*a) Trust evaluation with demand parameter level*

This level task is to compute the demand trust values and to select multiple servers as candidate of service provider. The parameters which are effective in service quality in this platform are: 1- delay 2- response time 3- accuracy 4- cost 5- availability 6- jitter. The user initialize mentioned parameters according to their importance in providing users request. Demand trust value will be calculated based on above parameters with accessing trust repository by weights middle method. In each computation $p$ servers that have maximum demand trust value will be selected as candidate of service provider. They will be transmitted to server selection and request allocation level. All of above processes are simulated by relation (4) until (8).

$$DP = (dp_1........dp_m). \quad (4)$$

$$w_i = \frac{dp_i}{\sum_{i=1}^{m} dp_i}, \quad \sum w_i = 1. \quad (5)$$

$$dtv_i = \sum_{i=1}^{n} \sum_{j=1}^{m} w_j, p_{j,i}. \quad (6)$$

TABLE I. SECURITY ATTRIBUTE

| | Security | Evaluation criteria | Security attribute |
|---|---|---|---|
| 1 | intrusion detection capability | Traffic audit data-size<br>Signature file size<br>Signature update frequency | $Sa_1$ |
| 2 | Antivirus capability | Memory scan frequency | $Sa_2$ |
| 3 | Firewall capability | Number of firewall rule | $Sa_3$ |
| 4 | Usage of secure network capability | TLS and/or IPsec | $Sa_4$ |
| 5 | Provision of execution sandbox | Isolated JVM | $Sa_5$ |
| 6 | Key management capability | Include Cryptographic function | $Sa_6$ |

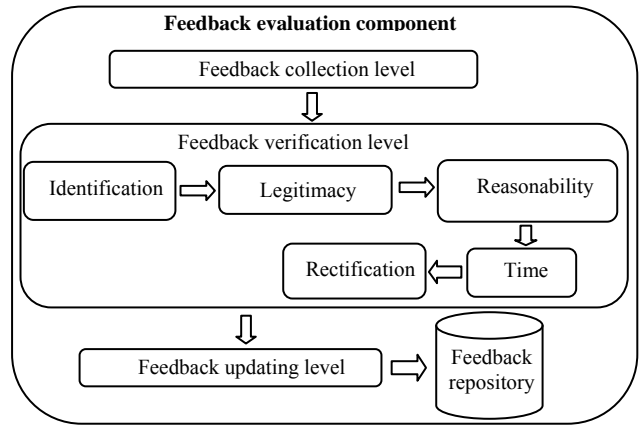

Figure 6. Feedback evaluation component

$$DTV_j = (dtv_1, dtv_2, dtv_3,......dtv_n). \quad (7)$$

$$DTV = \begin{bmatrix} DTV_1 \\ DTV_2 \\ . \\ . \\ . \\ DTV_k \end{bmatrix}. \quad (8)$$

In relation (4) *DP* is list of parameters initialized by user. In relation (5) $w_i$ represent weight of each parameter. In relation (6) $dtv_i$ stand for demand trust value of every service provider and m is the number of parameter. In relation (7) $dtv_i$ will be stored in $DTV_j$ for each request and n is the number of resource node. In relation (8) *DTV* represent an array of $DTV_j$ where as k is batch size.

*a) Server selection and request allocation level*

This level based on $DTV_j$ determined by relation (8) will select the appropriate service provider and will allocate the user request to selected service provider. This level has two sections. 1- Server selection based on roulette wheel mechanism 2- user request allocation.

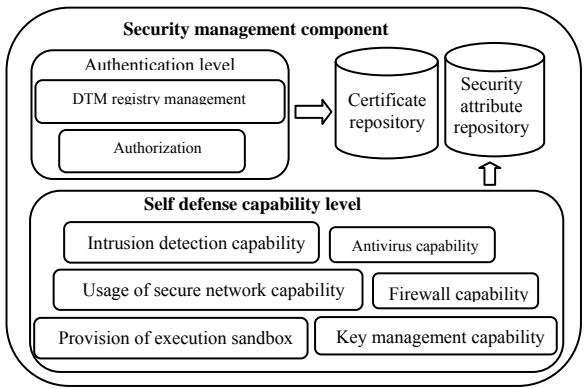

Figure 5. Security management component

```
1 feedback verification level (receive feedbacks from
feedback collection level)
2 Begin
3    for i=1 to m do // m is the number of parameter
4    $a = \frac{1}{l}\sum_{i=1}^{l} f_{pi}$
5    If  $(f_{pi}(new) - a > \delta)$   then
6        rectify ( )
7 feedback updating level ( )
8 End
```

Figure 7.  Feedback verification level

### SERVER SELECTION BASED ON ROULETTE WHEEL MECHANISM

This section uses roulette wheel mechanism to select appropriate service provider. The main reason of using this method is preserving load balance on all of service provider in a good manner. Relations (9) to (13) compute the percentage of user requests transmission to each service provider. In relation (9) *m* stand for the number of parameter and *w* is weight of each parameter. $P_i$ is the value of every parameter that has been stored in trust repository. In relation (10) *w* is received from relation (9). In relation (11) $dtv_i$ will be stored in *T.V* array. $sp_i$ represent the percentage of user sending request to *ith* service provider in relation (12). Finally in relation (13) $sp_i$ will be stored in *SP* array.

$$w = 1/m. \tag{9}$$

$$t.v_i = \sum_{i=1}^{n}\sum_{j=1}^{m} w \times p_{j,i} \tag{10}$$

$$T.V = (tv_1, tv_2, ........tv_n). \tag{11}$$

$$sp_i = \frac{tv_i}{\sum_{i=1}^{n} tv_i}. \tag{12}$$

$$SP = (sp_1, sp_2, .......sp_n), \quad \sum_{i=1}^{n} sp_i = 1. \tag{13}$$

### USER REQUEST ALLOCATION

This section allocates appropriate service provider between service provider candidates and appropriate service providers to user request by use of *SP* and roulette wheel mechanism.

*4) Trust evaluation component*

Trust evaluation component task is computing the servers trust values based on received feedback, users satisfaction value, domain self defense capability. Finally this component updates service provider trust value saved in trust repository. As shown in Fig.9 this component has two levels:
  a)  Trust value computing level
  b)  Trust value updating level.
  *a) Trust value computing level*

The task of this level is to calculate the user satisfaction value which it will be obtained from relation (14), whereas $P_{dmi}$ and $w_i$ have been received from demand trust evaluation component. $F'_{pi}$ is obtained from feedback evaluation component and *m* is the number of parameters described in demand trust evaluation component. Relation (15) computes the recommendation that $C_s$ is a number of successful recommendations and $C_f$ is a number of failed recommendations. In relation (16) *DF(new)* represent self defense capability which has been transmitted from security management component to trust evaluation component. Relation (17) will calculate trust value with using user satisfaction value, recommendation and self defense capability whereas *α, β* and *δ* are the weight of theirs.

$$S = \sum_{i=1}^{m} w_i \frac{|p_{dmi} - F'_{pi}|}{p_{dmi}}. \tag{14}$$

$$RE = \frac{c_s}{c_s + c_f}. \tag{15}$$

$$SD = DF(new). \tag{16}$$

$$T.V = \alpha.S + \beta.\text{Re} + \delta.SD, \alpha + \beta + \delta = 1. \tag{17}$$

*a) Trust value updating level*

The duty of Trust value updating level is updating trust repository with using below relation:

$$T_{new} = e^{-\beta.\Delta t}\frac{n}{n+1}T_{old} + (1 - e^{-\beta.\Delta t}\frac{n}{n+1})T.V. \tag{18}$$

Whereas $T_{new}$ represent new trust value, $T_{old}$ is old trust value, *N* stand for the current number of transaction, *T.V* is computed by relation (17) and *t* is the time difference between *T.V* and $T_{old}$. $e^{-\beta.\Delta t}$ represent a discount factor of $T_{old}$. Relation (18) is a reformed equation which earlier was used in [16] to calculate trust value. In last relation *T.V* has been computed from relation (17) whereas in [16] *r* was a trader's feedback.

*5) Accesses control component*

This component has the task of accesses control on available repository in lowest level of proposed platform.

*6) monitoring component*

Trust monitoring and trust re-evaluation is very important for implementation of TMS. Most of trust management solutions assume that trust is a static concept and therefore does not require monitoring or (periodic) re-evaluation. It involves updating or adding new information. as stated earlier the trust is dynamic in the real world as it changes with time. Trust monitoring ensures to reduce the risks involved [2].

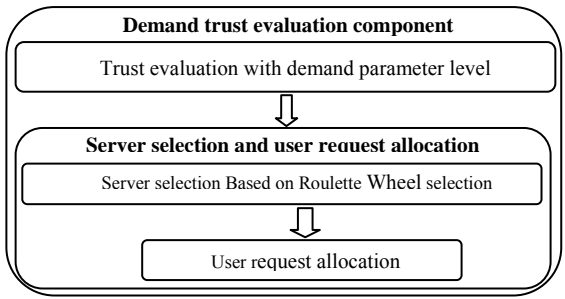

Figure 8. Demand trust evaluation component

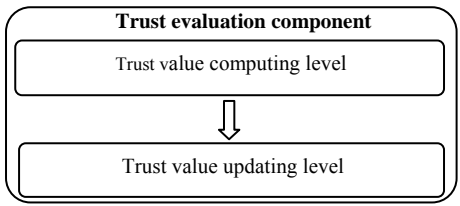

Figure 9. Trust evaluation component

IV. CONCLUSION AND FEATURE WORK

In this article the trust management systems with two levels in order to improving the security in grid systems has been proposed. In upper level there are trust negotiation, registration and initialization and propagation components which their tasks are adding new domain, registering and propagating new domain properties in grid systems. Also the lower level includes security management, feedback evaluation, demand trust evaluation, trust evaluation, access control and monitoring components. Their missions have been described in lower level section. The benefits of this platform are adding new domain in grid system, selecting one service provider which has closest adaption with user requests and using from domains security attribute as an important factor in computing the trust value. For future work we propose using of fuzzy method for computing trust value in trust evaluation component.

V. ACKNOWLEDGEMENT

This work was supported by Iran Telecommunication Research Center (ITRC).